\documentclass[fleqn,usenatbib]{mnras}

\usepackage{newtxtext,newtxmath}
\usepackage[T1]{fontenc}
\usepackage{ae,aecompl}

\usepackage{graphicx}	% Including figure files
\usepackage{amsmath}	% Advanced maths commands
\usepackage{amssymb}	% Extra maths symbols
\usepackage{subfig}
\usepackage{soul}

\title[Residual dispersion of MICADO]{Quantification of the expected residual dispersion of the MICADO Near-IR imaging instrument}

\author[J.A. van den Born et al.]{
J.A. van den Born,$^{1,2,3}$\thanks{E-mail: born@astro.rug.nl}
and W. Jellema$^{2,4}$
\\ \\
$^{1}$NOVA Optical Infrared Instrumentation Group at ASTRON, Oude Hoogeveensedijk 4, 7991 PD, Dwingeloo, The Netherlands\\
$^2$Kapteyn Astronomical Institute, University of Groningen, PO Box 800, 9700 AV Groningen, The Netherlands\\
$^{3}$Engineering and Technology Institute Groningen, University of Groningen, Nijenborgh 4, 9747 AG Groningen, The Netherlands\\
$^{4}$SRON Netherlands Insitute for Space Research, PO Box 800, 9700 AV Groningen, The Netherlands
}

\date{Accepted XXX. Received YYY; in original form ZZZ}

\pubyear{2020}

\begin{document}
\label{firstpage}
\pagerange{\pageref{firstpage}--\pageref{lastpage}}
\maketitle

\begin{abstract}
MICADO, a near-infrared imager for the Extremely Large Telescope, is being designed to deliver diffraction limited imaging and 50 micro arcsecond (\textmu as) astrometric accuracy. MICADO employs an atmospheric dispersion corrector (ADC) to keep the chromatic elongation of the point spread function (PSF) under control. We must understand the dispersion and residuals after correction to reach the optimum performance. Therefore, we identified several sources of chromatic dispersion that need to be considered for the MICADO ADC. First, we compared common models of atmospheric dispersion to investigate whether these models remain suitable for MICADO. We showed that the differential dispersion between common atmospheric models and integration over the full atmosphere is less than 10 \textmu as for most observations in \textit{H}-band. We then performed an error propagation analysis to understand the uncertainty in the atmospheric dispersion as a function of atmospheric conditions. In addition, we investigated the impact of photometric color on the astrometric performance. While the differential refraction between stars within the same field of view can besignificant, the inclusion of an ADC rendered this effect negligible. For MICADO specifically, we found that the current optomechanical design dominates the residual dispersion budget of 0.4 milli arcseconds (mas), with a contribution of 0.31 mas due to the positioning accuracy of the prisms and up to 0.15 mas due to a mismatch between the dispersive properties of the glass and the atmosphere. We found no showstoppers in the design of the MICADO ADC for achieving 50 \textmu as relative astrometric accuracy.
\end{abstract}


%%%%%%%%%%%%%%%%%%%%%%%%%%%%%%%%%%%%%%%%%%%%%%%%%%

%%%%%%%%%%%%%%%%% BODY OF PAPER %%%%%%%%%%%%%%%%%%

\section{Introduction}
The next generation of large telescopes, such as the Extremly Large Telescope (ELT) \citep{ESOconstrprop}, the Thirty Meter Telescope \citep{Sanders2014} and the Giant Magellan Telescope \citep{Johns2012}, offer a significant increase in aperture diameter. With this increase in telescope size, several unwanted optical effects become important or can no longer be assumed negligble and have to be reconsidered (e.g. \cite{Devaney2008,Jolissaint2010,Trippe2010,Ellerbroek2013}). This directly follows from the relation between the angular size of the point spread function (PSF), $\theta_{\textnormal{PSF}}$, the telescope diameter, $D$, and the wavelength, $\lambda$.
\begin{equation}
    \theta_{\textnormal{PSF}} = 1.22 \frac{\lambda}{D}\label{eq:airyPattern}.
\end{equation}
 
Besides an increase in resolution, these large telescopes will offer capabilities for high precision relative astrometry, allowing astronomers to measure relative angular distances between stars to several tens of micro arcseconds \citep{Trippe2010,Schoeck2013,Massari2016}. Consequently, the precise PSF shape and position must be understood to a new level of precision.

The near infrared instrument offering such astrometric accuracy on the ELT will be MICADO, the Multi-Adaptive Optics Imaging CamerA for Deep Observations \citep{Davies2016}. It will offer relative astrometric accuracy of 50 micro arcseconds (\textmu as). In order to reach this level of performance it is essential to correct for various atmospheric effects. The most prevalent of these, atmospheric turbulence, is taken care of by the adaptive optics systems accompanying MICADO. The next most important effect is the chromatic dispersion that increases as the telescope moves away from zenith. Due to the wavelength dependence of the atmosphere's refractive index, light with shorter wavelengths is refracted more than light with longer wavelengths, causing an elongation of the PSF. To counteract this effect, MICADO incorporates an atmospheric dispersion corrector (ADC) consisting of two mirrored counter-rotating Amici prisms. The PSF elongation can be controlled by rotating these prism sets. Although most of the atmospheric dispersion can be reversed this way, some residual dispersion will inevitably persist.

Because MICADO will be a diffraction limited instrument, the residual dispersion in its imaging mode is required to be kept under 2.5 milli arcseconds (mas) in \textit{J}, \textit{H} and \textit{K}-band, with a stated goal of 1 mas. However, the root-mean-square stability of the PSF during a two minute observation should not exceed 0.4 mas in order to reach the expected astrometric performance of 50 \textmu as \citep{Pott2018}. Therefore a residual dispersion less than 0.4 mas is considered the leading requirement for the design of the MICADO ADC. The analyses in this work will be presented for \textit{H}-band.

A proper understanding of all dispersive effects should first be established in order to minimize the residual dispersion and to decide on the best control algorithm of an ADC. \cite{Spano2014} has shown that there are considerable differences between various refractivity models used in the calculation of atmospheric refraction. In complement to this analysis, we compare various atmospheric models to assess how these impact the expected refraction. Furthermore, while other works have studied atmospheric dispersion in considerable detail (e.g. \cite{Danjon1980,Gubler1998,Mangum2015,Corbard2019}), much less is written about how this relates to the performance of an ADC employed at an extremely large telescope. By finding a sufficiently detailed model of the atmosphere and ADC, we're able to study the various systematic and random contributors to the chromatic dispersion expected on the image plane of MICADO. We will first review several methods to calculate atmospheric dispersion in section~\ref{sec:AtmDispersion}. Next, we will describe a mathematical model of an ADC and how it can be configured to reverse the atmospheric dispersion in section~\ref{sec:ADC}. In section~\ref{sec:results} we compare the amount of atmospheric dispersion given by the different models. Then, we propose and quantify several other factors that impact the residual dispersion in the central part of the image plane, either due to random measurement errors or due to systematic limitations of the instrument design. Finally, we shortly discuss some other related considerations in section~\ref{sec:discussion} and present our conclusions in section~\ref{sec:conclusion}.

\section{Atmospheric dispersion}\label{sec:AtmDispersion}
\begin{figure}
    \centering
    \includegraphics[width=0.8\linewidth]{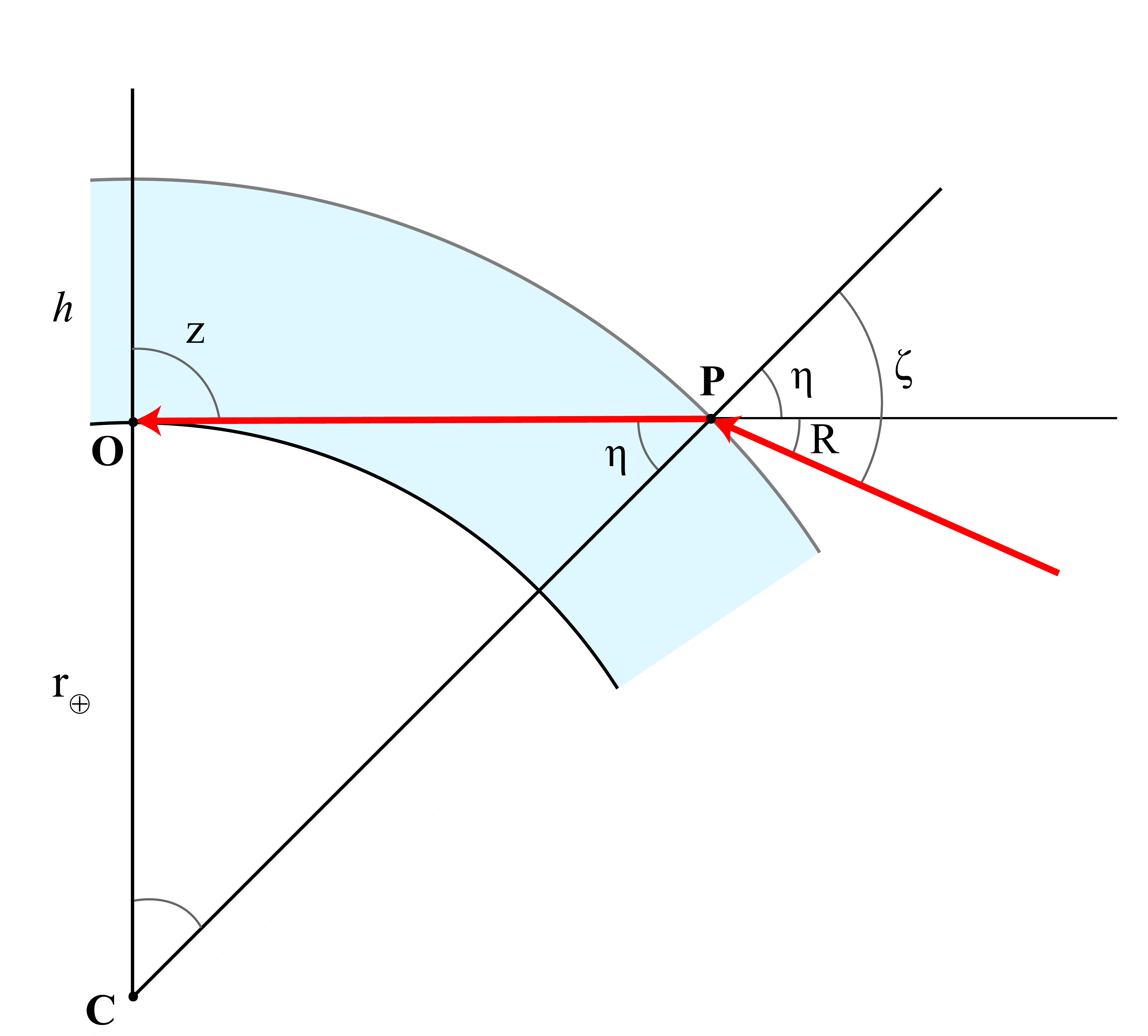}
    \caption{Diagram illustrating refraction in a homogeneous spherical atmosphere. The radius of curvature of the earth is given by $r_{\oplus}$, the height of the atmosphere is $h$ and the incoming ray has a local zenith distance $\zeta$, before it gets refracted and observed at $O$ at a zenith distance $z$. The angle of the ray after refraction at $P$ is denoted by $\eta$.}
    \label{fig:sphericalAtmosphere}
\end{figure}

Atmospheric dispersion is defined as the difference in the angle of refraction between two light rays of different wavelength after passing through the atmosphere. The wavelength dependency only originates from the used refractivity model, although the overall dispersion may change as a function of the chosen geometry and atmospheric conditions.

In this section we consider a plane-parallel model and various spherical shell models, for which we adopt the geometry given in Figure~\ref{fig:sphericalAtmosphere}, to discuss the most common ways of calculating the refraction of light passing through our atmosphere.

\subsection{Atmospheric models}\label{sec:AtmosphereMod}

\subsubsection{Plane-parallel atmosphere model}
The simplest model assumes a homogeneous plane-parallel atmosphere, corresponding in Figure~\ref{fig:sphericalAtmosphere} to the limit where $z\approx\eta$. This model is valid for small zenith angles only, as the underlying assumption grows erroneous with increasing zenith distance. The refraction in this model is derived from Snell's law. 
\begin{align}
    R &= \zeta - z \nonumber\\
        &= \sin^{-1}({n\sin z}) - z \nonumber\\
        &\approx (n-1) \tan z, \label{eq:refPP}
\end{align}
where the refraction $R$ is the difference between the observed zenith angle, $z$ and the local zenith angle, $\zeta$, before the ray gets refracted and $n$ is the refractive index of the atmospheric air. For small angles, the atmospheric dispersion is then described by
\begin{equation}
    \Delta R = R(\lambda_1) - R(\lambda_2) = (n_1 - n_2) \tan z. \label{eq:flatEarthRefDisp}
\end{equation}

\subsubsection{Cassini's refraction model}\label{sec:cassiniModel}
Assuming a spherical and homogeneous atmosphere by releasing the $z\approx\eta$ constraint of the plane-parallel atmosphere, a surprisingly accurate approximation for the refraction can be written as
\begin{equation}
    R = \zeta - \eta = \sin^{-1}\bigg( \frac{nr_{\oplus}\sin z}{r_{\oplus} + h} \bigg) - \sin^{-1}\bigg(\frac{r_{\oplus}\sin z}{r_{\oplus} + h} \bigg) \label{eq:cassini}.
\end{equation}
Here, $r_{\oplus}$ denotes the radius of the earth and $h$ is the height of the atmosphere.
This description of refraction is often referred to as \textit{Cassini's homogeneous refraction model} \citep{Young2006}.

In a homogeneous atmosphere only the density and pressure at the observer need to be known to derive the total atmospheric refraction. We therefore use the adiabatic scale height, or \textit{reduced height},
\begin{equation}
    h_r = \frac{p_o}{\rho_o g}, \label{eq:adiabaticScaleHeight}
\end{equation}
where $p_o$ and $\rho_o$ are the pressure and atmospheric density at the observer and $g$ is the gravitational acceleration.

Alternatively, we could assume an isothermal atmosphere to find the isothermal scale height
\begin{equation}
    h_r = \frac{k_b T_o}{mg}, \label{eq:isoScale}
\end{equation}
where $k_b$ is the Boltzmann constant, $T_o$ is the temperature at the observer and $m$ is the mass in kilograms of an average molecule of air. Both scale heights are around 8 kilometers at typical atmospheric conditions.

For improved accuracy, we also use the locally experienced gravity and the local curvature of the Earth from the Geodetic Reference System of 1980 \citep{HeiskanenMoritz1967, GRS80}, in the above equations.

\subsubsection{Refraction integral}\label{subsec:full_integration}
Without making assumptions about the atmosphere, full integration over the optical path through the atmosphere is required to find the observed refraction. Analogous to the $n\sin\zeta$ invariant for a plane parallel geometry, the refractive invariant for a spherical atmosphere, $nr\sin \zeta$, is constant as function of distance from the center of the Earth, $r$. From this the refraction integral can be derived (e.g. \cite{Young2006}),
\begin{equation}
    R = \int^{n_o}_{1}\tan(\zeta)\frac{dn}{n}. \label{eq:refractionIntegral}
\end{equation}
Here $n_o$ is the refractive index of the air at the observer. 

The refraction integral allows a full numerical integration of the path of the light ray through the atmosphere. However, it requires the conditions at each point along that path to be known.

\cite{Auer2000} have shown that it is possible rewrite equation~\ref{eq:refractionIntegral} in a way that prevents the refraction from going to infinity for large zenith angles. It is also easier to evaluate numerically.
\begin{equation}
    R = - \int^{z}_{0} \frac{d(\ln{n})/d(\ln{r})}{1+d(\ln{n})/d(\ln{r})}d\zeta. \label{eq:auerInt}
\end{equation}

\subsubsection{Error function model}
\cite{Corbard2019} discuss a different model of the form of a Gauss error function, which had initially been found by \cite{Fletcher1931} and was also derived by \cite{Danjon1980}. This model assumes an exponential decrease in the atmospheric density as a function of height. It can be shown that the refraction can then be described by
\begin{equation}
    R = \alpha \bigg( \frac{2 - \alpha}{\sqrt{2\beta-\alpha}}\bigg) \sin(z) \Psi\bigg( \frac{\cos(z)}{\sqrt{2\beta-\alpha}} \bigg), \label{eq:erfRef}
\end{equation}
where $\alpha=n-1$ is the local air refractivity and $\beta=h_r/r_o$ is the ratio between the reduced height of the atmosphere, equation~(\ref{eq:adiabaticScaleHeight}) or (\ref{eq:isoScale}), and the radius of curvature of the Earth at the observer. The function $\Psi(x)$ in the equation above is defined as
\begin{equation}
    \Psi(x) = e^{x^2}\int^{\infty}_{x}e^{-t^2}dt = \frac{\sqrt{\pi}}{2}e^{x^2}\big(1-\textnormal{erf}(x)\big).
\end{equation}

\subsubsection{Oriani's theorem}
Probably the most well known form of an atmospheric refraction formula is the formula first derived by Oriani in the eighteenth century \citep{Oriani1787}, of the form
\begin{equation}
    R = A \tan(z) + B \tan^{3}(z), \label{eq:tan3model}
\end{equation}
with $A$ and $B$ being constants. To describe $A$ and $B$ analytically, a Laurent series expansion can be done on $\Psi(x)$ in equation~(\ref{eq:erfRef}). Then we find the following expression for the atmospheric refraction.
\begin{equation}
R = \alpha (1 - \beta) \tan (z) - \alpha \Big( \beta - \frac{\alpha}{2}\Big) \tan^{3}(z) + 3\alpha \Big(\beta - \frac{\alpha}{2}\Big)^2\tan^5(z),  \label{eq:tan5model}  
\end{equation}
where the same definitions for $\alpha$ and $\beta$ have been used as for the error function model, discussed above. Aside from the first and third order coefficients present in equation~(\ref{eq:tan3model}), equation~(\ref{eq:tan5model}) also includes a fifth order term.

\subsubsection{Our preferred model}
\begin{figure}
    \includegraphics[width=\columnwidth]{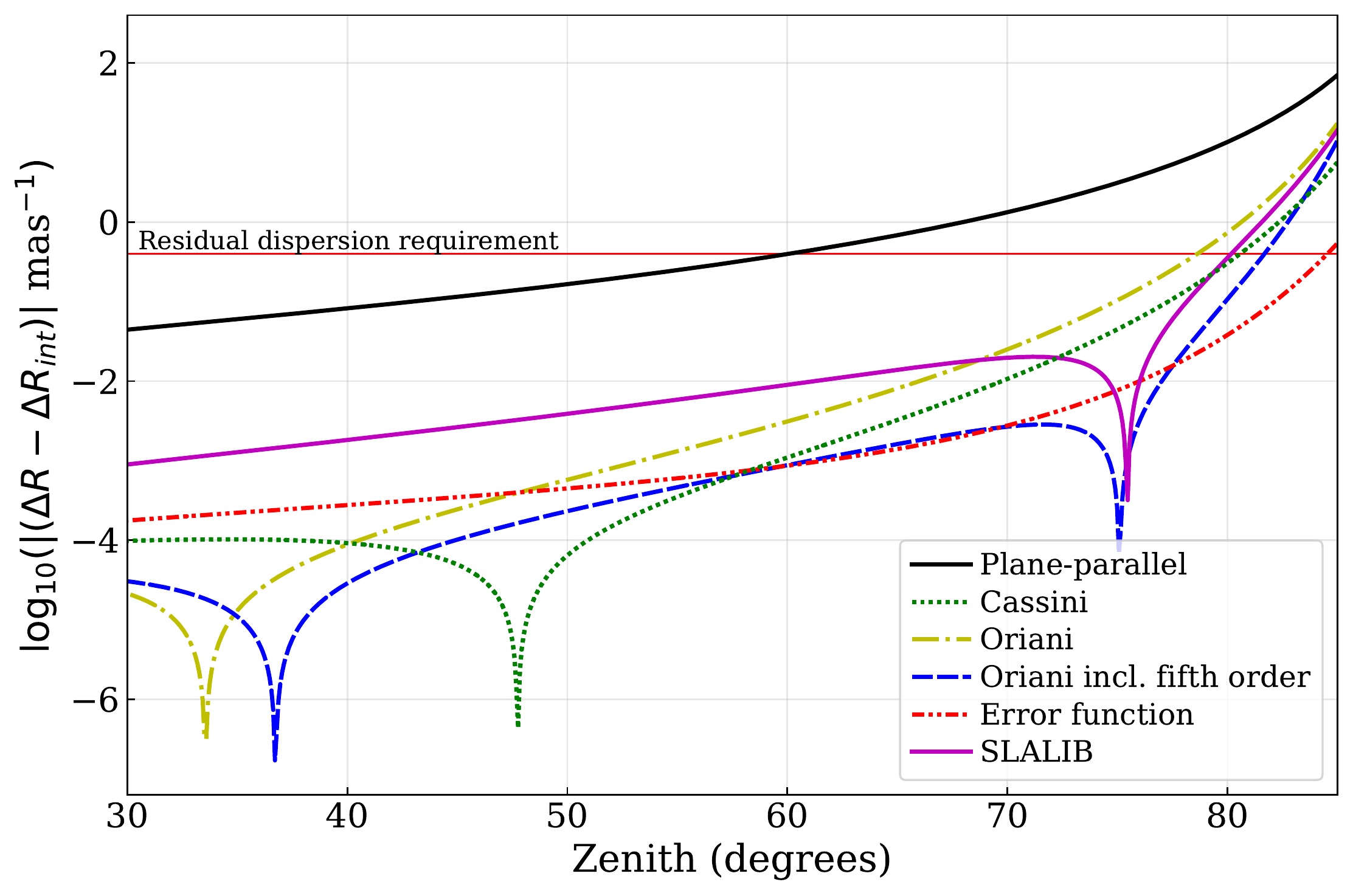}
    \caption{The atmospheric dispersion in \textit{H}-band in milli arcseconds relative to the full integration method for the different refraction models. Within the observational range of the MICADO all the models, except the plane-parallel atmosphere model, agree to within 10 \textmu as. This is a small fraction of the 0.4 mas allowed residual dispersion.}
    \label{fig:dispersionModels}
\end{figure}

\begin{table*}
    \centering
    \caption{Refraction for $\lambda=1.49$ \textmu m in arc seconds as a function of zenith angle for the different refraction models for standard atmospheric conditions at sea level ($T=288.15$ K, $p=101325$ Pa, $H=0.0$ and $x_c=314$ ppm). The adiabatic scale height is used for the calculations.}
    \begin{tabular}{rrrrrrr}
    \hline
    $z$ & Plane-parallel & Cassini & Oriani (3$^{\textnormal{rd}}$) & Oriani (5$^{\textnormal{th}}$) & Error function & Full integration \\
  \hline
 10\degr &   9.940 &    9.926 &    9.926 &    9.926 &    9.926 &     9.926 \\
 30\degr &  32.547 &   32.489 &   32.489 &   32.489 &   32.490 &    32.489 \\
 50\degr &  67.193 &   66.977 &   66.977 &   66.977 &   66.978 &    66.978 \\
 60\degr &  97.677 &   97.160 &   97.157 &   97.161 &   97.164 &    97.162 \\
 70\degr & 155.038 &  153.298 &  153.274 &  153.312 &  153.321 &   153.314 \\
 80\degr & 321.112 &  307.688 &  306.982 &  308.399 &  308.253 &   308.142 \\
 85\degr & 656.255 &  561.900 &  542.875 &  589.985 &  571.503 &   570.005 \\
\hline
\end{tabular}    
    \label{tab:refractionComparison}
\end{table*}

\begin{table*}
    \centering
    \caption{Dispersion for \textit{H}-band (1.49 \textmu m to 1.78 \textmu m) in milli arc seconds as a function of zenith angle for the different refraction models. Standard atmospheric conditions at sea level ($T=288.15$ K, $p=101325$ Pa, $H=0.0$ and $x_c=314$ ppm) are assumed. Again, we've used the adiabatic scale height for our calculations.}
    \begin{tabular}{rrrrrrr}
    \hline
    $z$ & Plane-parallel & Cassini & Oriani (3$^{\textnormal{rd}}$) & Oriani (5$^{\textnormal{th}}$) & Error function & Full integration \\
    \hline
10\degr &     7.565 &    7.554 &    7.554 &    7.554 &     7.554 &     7.554 \\
30\degr &    24.771 &   24.727 &   24.727 &   24.727 &    24.727 &    24.727 \\
50\degr &    51.147 &   50.982 &   50.982 &   50.982 &    50.983 &    50.983 \\
60\degr &    74.367 &   73.973 &   73.971 &   73.973 &    73.975 &    73.974 \\
70\degr &   118.112 &  116.784 &  116.770 &  116.792 &   116.797 &   116.794 \\
80\degr &   245.468 &   235.12 &  234.695 &  235.526 &   235.459 &   235.421 \\
85\degr &   508.856 &  433.512 &  421.898 &  449.557 &   439.625 &   439.090 \\
    \hline
    \end{tabular}
    \label{tab:dispersionComparison}
\end{table*}

Unless stated otherwise, we use Cassini's refraction model to describe atmospheric dispersion. We justify this choice by assessing the agreemement between the different models for zenith angles below 85$^{\circ}$. We use as a reference the full integration method of section~\ref{subsec:full_integration} over the U.S. Standard Atmsophere \citep{1976USSA}. This data set gives us the temperature and pressure profiles up to an altitude of 86 kilometers. We assume that the refraction above this point is negligible.

Tables~\ref{tab:refractionComparison} and \ref{tab:dispersionComparison} show several refraction and dispersion values for the discussed refraction models. Since the ELT will not be observing any lower than about 25 degrees of elevation, the differences between the various models with respect to the atmospheric dispersion for H-band will be minimal compared to the 0.4 mas residual dispersion requirement. This is illustrated in Figure~\ref{fig:dispersionModels}, where the dispersion of the various atmospheric dispersion models is plotted relative to the full integration method as a function of zenith angle, using the same refractivity for all models. The singularities in log-space of Figure~\ref{fig:dispersionModels} correspond to the zenith angle at which the dispersion curve of the respective model crosses the dispersion curve of the full integration model. Note that the simple expression of the Cassini homogeneous atmosphere model gives accurate results up to 65$^{\circ}$ of zenith. Overall, none of the spherical atmosphere models differs by more than 10 \textmu as for zenith angles less than 60$^{\circ}$.

Because Cassini's refraction model is mathematically simple, the complexity of mathematical operations decreases significantly without much loss of accuracy. This makes it our preferred model for this work.

\subsection{Refractivity model}
An accurate description of the refractive index of atmospheric air is necessary to describe the atmospheric dispersion accurately. \cite{Spano2014} compared the most common refractivity models to assess the accuracy of the atmospheric surface model included in the optical design software Zemax OpticStudio. The author confirmed that OpticStudio used an outdated equation from \cite{BS1939}. Compared to the more recent work of \cite{Ciddor1996} a difference in atmospheric dispersion of 0.8 milli arcseconds (mas) can be present in \textit{I}-band (800 to 934 nm) for a moderate zenith distance of 30\degr. If we compare the results from \citeauthor{Ciddor1996} with other measurements, such as reported in \cite{Birch1993} or \cite{Bonsch1998}, then the differential is reduced to less than 0.2 mas under the same conditions. 
We have selected the \citeauthor{Ciddor1996} refractivity model for this work, because it is the standard equation of International Association of Geodesy (IAG) for the calculation of the index of refraction of atmospheric air. The model is expected to be accurate over a wide range of temperature, pressure and humidity levels. It is valid for most of the wavelength coverage of MICADO, from \textit{I}-band up to the majority of \textit{H}-band. For \textit{K}-band and longer wavelengths the absorption lines of OH and H$_2$O vapour start to impact the refractive index, in which case the model by \cite{Mathar2007} could be preferred. The validity of this model has been tested on-sky in the mid-infrared by \cite{Skemer2009}, making it the preferred model at those wavelengths.

The fundamental assumption for most refractivity models is that the index of refraction, $n$, of atmospheric air scales with the density $\rho$, either via the Gladstone-Dale relation, $(n-1)\propto\rho$, or through the Lorentz-Lorenz relation, $(n^2-1)/(n^2+2)\propto\rho$ \citep{Ciddor1996,Kragh2018}. We use the latter. By scaling $n$ and $\rho$ to a set of reference conditions, the refractive index for the atmospheric conditions of interest is computed.

Similar to \citeauthor{Ciddor1996}, we use the CIPM-81/91 equation~\citep{Davis1992} to calculate the atmospheric density. A revised equation was published in \cite{Picard2008}. This updated equation results in a slight increase of the atmospheric density, but it does not result in any significant change in the refractive index. Because of this and for validation purposes, we have adopted the older CIPM-81/91 equation for this work.

The combination of the density equations and the reference refractivity measurements allows us to compute the refractivity of atmospheric air, and therefore also the atmospheric refraction, as a function of wavelength $\lambda$ in micrometers (\textmu m), temperature $T$ in Kelvin (K), pressure $p$ in Pascal (Pa), relative humidity $H$ and the CO$_\textnormal{2}$ density $x_c$ in parts per million (ppm).

%%%%%%%%%%%%% Section 3 %%%%%%%%%%%%%%%%%
\section{Description of the ADC}\label{sec:ADC}
The Helmholtz reciprocity principle states that the chromatic dispersion of light must be reversible. Correspondingly, in this section, we'll explain how to reverse the undersired effect of atmospheric dispersion by configuring the dispersive properties of a set of multiple prisms. We consider here the counter-rotating atmospheric dispersion corrector, where two prisms rotate away from each other to control the amount of dispersion.

\subsection{Dispersion of an atmospheric dispersion corrector}
\begin{figure}
    \centering
    \includegraphics[width=\linewidth]{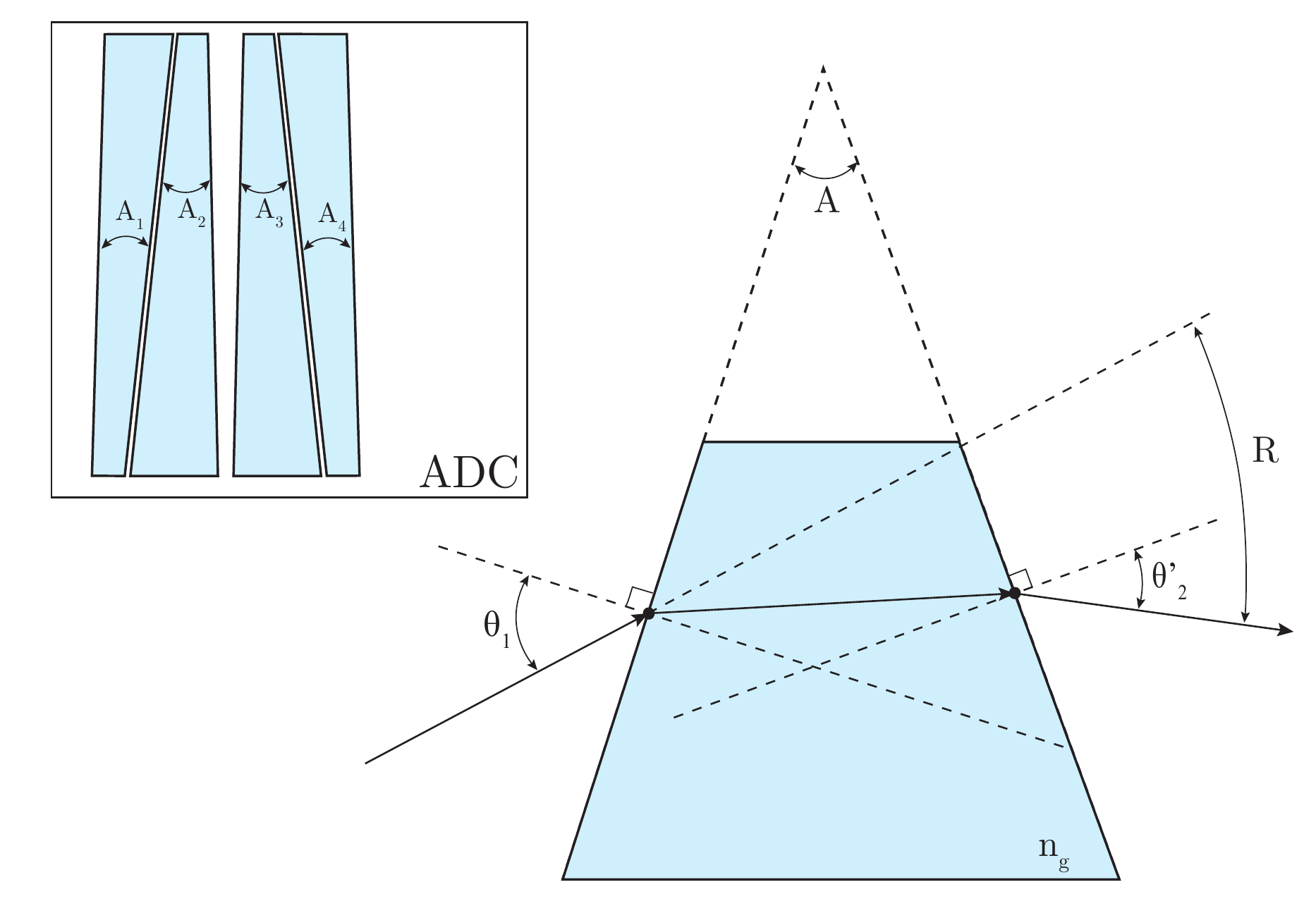}
    \caption{Geometry of a prism. The insert shows the geometry of a typical atmospheric dispersion corrector, shown in it's maximum dispersion correction position. For the MICADO ADC, $A_1=A_4$ and $A_2=A_3$. Also, $A_1$ and $A_2$ have opposite sign.}
    \label{fig:prism_schematic}
\end{figure}

We consider a single prism, with a geometry as illustrated in Figure~\ref{fig:prism_schematic}. The deviation of a light ray passing through this prism, denoted by $R$, can be calculated using Snell's law at each glass interface. For small angles and assuming that the prism is located inside a vacuum, this deviation is described by
\begin{equation}
    R = (n_g - 1)A, \label{eq:prismDeviationApprox}
\end{equation}
where $n_g$ is the refractive index of the glass at the wavelength of interest and $A$ is the apex angle of the prism. For a complete description of prism refraction, see \cite{Hagen2011}.

In the case of $N$ sequential prisms, the total deviation of the ray can be calculated by summation of the ray deviation from each respective prism, taking into account that the angle of incidence for each prism is different. Assuming the ADC is located in the pupil plane and etendue is conserved, we may scale the angular deviation at the ADC to the angular deviation in the sky by a multiplication of the ratio between the ADC diameter, $D_{\textnormal{\tiny{ADC}}}$, and the entrance pupil diameter, $D_{\textnormal{\tiny{EPD}}}$. Our ADC model is thus,

\begin{equation}
    R_{\textnormal{sky}} = \frac{D_{\textnormal{\tiny{ADC}}}}{D_{\textnormal{\tiny{EPD}}}}\sum_{i=1}^N R_i. \label{eq:adc_to_sky_refraction}
\end{equation}

Atmospheric dispersion is predominantly pointed away from zenith and therefore the ADC must be configured such that it inverses the atmospheric dispersion in that direction. This is done by counter-rotating the two prisms, effectively changing the apex angles of the prisms.

The change in the apex angle is found by considering a rotation matrix in a fixed cartesian coordinate frame. We consider the apex angles in the $x$ and $y$ directions.
\begin{equation}
    \begin{bmatrix}
    A'_x \\ A'_y
    \end{bmatrix} = 
    \begin{bmatrix} 
        \cos \theta & -\sin\theta \\
        \sin \theta & \cos \theta
    \end{bmatrix} 
    \begin{bmatrix}
    A_x \\ A_y
    \end{bmatrix}\label{eq:tipTilt}
\end{equation}

The optimum position of the ADC is found when the prisms are rotated with respect to each other such that the dispersion of the ADC equals that of the atmosphere. Applying the rotation matrix of equation~(\ref{eq:tipTilt}) to each of the ADC prisms to find the corresponding apex angles and then using equation~(\ref{eq:prismDeviationApprox}) gives us an approximation of the refraction through a prism. We assume that equal but opposite rotation $\theta$ takes place for both Amici prisms and that the prisms have equal apex angles and refractive indices. We also assume that rotation starts from the maximum dispersive configuration, so that we may neglect the $y$ component of the initial apex angles, as shown in Figure~\ref{fig:prism_schematic}. Hence, we use scalar notation and not vector notation in the following. Then we find the refraction through the ADC as
\begin{equation}
    R_{\textnormal{\tiny{ADC}}} = (n_g - 1)(A_1 + A_2)\cos\theta
\end{equation}
and the dispersion as
\begin{equation}
    \Delta R_{\textnormal{\tiny{ADC}}} = \Delta n_g (A_1 + A_2)\cos\theta. \label{eq:adc_dispersion_approx}
\end{equation}
Here $\Delta n_g$ is the differential refractive index for the two wavelengths of interest.

Combining the plane parallel dispersion model, equation~(\ref{eq:flatEarthRefDisp}), with the ADC model, equations~(\ref{eq:adc_to_sky_refraction}) and (\ref{eq:adc_dispersion_approx}), we find an approximation of the optimum rotation angle as a function of zenith distance,
\begin{equation}
    \theta_{\textnormal{opt}} = \cos^{-1}\big(c_f \tan z\big), \label{eq:optPos}
\end{equation}
where $\theta_{\textnormal{opt}}$ is the rotation of the prism in counter clockwise direction for the top prism and clockwise direction for the bottom prism. The filter constant $c_f$ has a specific value for each wavelength band and can be written as
\begin{equation}
    c_f = \frac{D_{\textnormal{\tiny{EPD}}}}{D_{\textnormal{\tiny{ADC}}}} \frac{\Delta n_{\textnormal{atm}}}{\Delta n_g (A_1 + A_2)}. \label{eq:filter_constant}
\end{equation}

Due to the assumption of a plane parallel atmosphere, the error in $\theta_{\textnormal{opt}}$ grows unacceptably large for an ELT at moderate to large zenith angles. In the analyses where an accurate positioning is preferred, we apply a simple iterative optimization algorithm to find $\theta_{\textnormal{opt}}$ for the ADC model. For smaller telescopes, however, it is an effective model for the positioning of an ADC \citep{Egner2010}. Furthermore, equation~(\ref{eq:optPos}) will prove useful in our analysis of the necessary rotation resolution (sections~\ref{sec:mechTol} and \ref{subsec:AtmDispersion/mech}), where we only need to consider small zenith angles.

\subsection{Discretization of the dispersion correction}\label{sec:mechTol}
The present design of the MICADO ADC makes use of stepper motors to rotate the prisms. Inherent to this design choice is a finite positioning resolution and consequently a limit to the dispersion correction accuracy. Here we calculate the necessary ADC rotation per unit of dispersion, for a given zenith angle. This will allow us to find the required number of steps to achieve a residual dispersion less than 0.4 mas. First, we note that the atmospheric dispersion per degree zenith angle is the derivative of the atmospheric dispersion formula. Differentiating the Cassini atmospheric model of equation~(\ref{eq:cassini}) with respect to the zenith angle, we find
\begin{equation}
    \frac{d(\Delta R)}{dz} = \frac{n_1 \mathcal{R} \cos{z}}{\sqrt{1-n_1^2 \mathcal{R}^2 \sin^2z}} - \frac{n_2 \mathcal{R} \cos{z}}{\sqrt{1-n_2^2 \mathcal{R}^2 \sin^2z}}\label{eq:dispersionderivative},
\end{equation}
where we have taken $\mathcal{R} = r_{\oplus} / (r_{\oplus} + h_r)$.

The differentation of the optimum rotation angle, equation~(\ref{eq:optPos}), with respect to the zenith angle is 
\begin{equation}
    \frac{d\theta_{\textnormal{opt}}}{dz} = \frac{c_f \sec^2 z}{\sqrt{1 - c_f^2 \tan^2 z}}.\label{eq:optPosderivative}
\end{equation}
The ADC rotation per unit of dispersion is then the ratio of equation~(\ref{eq:optPosderivative}) over equation~(\ref{eq:dispersionderivative}). We find that the ADC rotation per unit of dispersion is smallest for small $z$. In other words, the number of steps the ADC must be able to do is determined by the smallest zenith angle for which the requirement on dispersion correction is defined.

\section{Results}\label{sec:results}
\subsection{Systematic errors of the atmospheric dispersion calculation}\label{subsec:results_and_errors}

The established framework is now used to determine the systematic errors of the atmospheric dispersion calculations. In section~\ref{sec:AtmosphereMod} we found that changing the atmospheric model will result in differential atmospheric dispersion less than 10 \textmu as for observations done by MICADO.

Larger errors in the atmospheric dispersion result from variations in the atmospheric conditions. For example, in H-band under standard atmospheric conditions at sea level ($T=288.15$ K, $p=101325$ Pa, $H=0.0$ and $x_c=314$ ppm) a change in temperature of 1 K changes the refraction by roughly 0.1 arcsec and the dispersion by roughly 0.1 mas. The pressure and relative humidity variations have less impact, but should be included as well. Presently, the MICADO consortium is still considering how to implement corrections for these parameter dependencies, possibly in the form of a lookup table or by including the model described in this work.

Error propagation was performed on the Ciddor--Cassini refraction model to assess the impact of measurement uncertainties, assuming they are uncorrelated, using
\begin{align}
     \sigma_{\Delta R}^2 &= \sum_{x\in S} \Big(\frac{\partial \Delta R}{\partial x}\sigma_{x}\Big)^2, && \textnormal{where } S=\{p,T,H,x_c,\lambda,z\}.\label{eq:dispError}
\end{align}
Here $\sigma_x$ denotes the uncertainty in $x$, for $x\in\{p,T,H,x_c,\lambda,z\}$. Monte Carlo simulations were used to verify our calculations of equation~\ref{eq:dispError}. A comparison of the results for some realistic measurement uncertainties is shown in Table~\ref{table:dispersionTable}. An exploration of a large part of the parameter space of the uncertainties is given in Appendix~\ref{appendix:uncertainties}. 

Our analysis shows that realistic uncertainties in $T$, $p$ and $H$ do not have a very significant impact on the dispersion uncertainties. It will not be necessary to measure the CO$_2$ densities at the observing site, because the uncertainty in $x_c$ impacts the atmospheric dispersion error only on the order of nano arcseconds. The pointing error of the telescope will be some tenths of arcseconds. For example, on VLT UT2 the tracking error is around 0.1 arcseconds, after the closed loop tracking system has locked on a guide star \citep{Nurzia2018}. We find that the dispersion error contribution of an uncertainty in the zenith angle will be negligible if we assume similar performance for the ELT. In contrast to the other parameters, the uncertainty in the wavelength can easily dominate the overall dispersion uncertainty.

We illustrate this by taking the measurement uncertainties of the VLT Astronomical Site Monitor, where $\sigma_p=10$ Pa, $\sigma_T=0.2^{\circ}$C, $\sigma_H=0.01$ and $\sigma_{x_c}=20$ ppm \citep{VLTASM}. Neglecting the wavelength dependency, we can expect a systematic uncertainty of 10 to 20 \textmu as in \textit{H}-band at $z=30^{\circ}$, as shown in the third line of Table~\ref{table:dispersionTable}. If we assume an uncertainty of 1~nanometer for the wavelength, then this uncertainty dominates over the other variables by nearly an order of magnitude and becomes a non-negligible fraction of the MICADO requirement on dispersion correction.

This last result implies that the cut-on and cut-off wavelengths should be defined carefully and that the uncertainty should be minimized through accurate measurements of the bandpass filters.

\begin{table}
\centering
\caption{The atmospheric dispersion uncertainty calculated for \textit{H}-band, at a zenith distance of 30$^{\circ}$ at standard atmospheric conditions. In the fourth column a standard uncertainty in the wavelength $\sigma_{\lambda}=1$ nm is assumed, except on the first line. No uncertainty in the zenith distance is assumed. The third line takes the uncertainties from the VLT Site Monitor \citep{VLTASM}.}\label{table:dispersionTable}
\begin{tabular}{rrrrc}
\hline
$\sigma_p$ & $\sigma_T$ & $\sigma_H$ & $\sigma_{\Delta R}$ & $\sigma_{\Delta R} (\sigma_{\lambda} = 0)$ \\ 
(Pa) & (K) & & (\textmu as) & (\textmu as) \\ \hline
0   & 0.0 & 0.00 & 0.0    & 0.0\\
10  & 0.1 & 0.01 & 129.0  & 9.1\\
10  & 0.2 & 0.01 & 129.8  & 17.5\\
20  & 0.2 & 0.02 & 129.9  & 18.1\\
100 & 1.0 & 0.05 & 156.9  & 89.8\\
\hline
\end{tabular}
\end{table}

\subsection{The Zemax atmospheric dispersion model}
The atmospheric dispersion model by \cite{H&S1985} is used as the atmosphere model of Zemax OpticStudio and for refraction and dispersion calculations in the popular SLALIB library \citep{SLALIB}. 

This model assumes a two layer atmosphere, consisting of a troposphere with a constant decrease in temperature below 10 km and a stratosphere with a constant temperature up to a height where the refraction can be considered negligble, usually at 80 km. The pressure and temperature profiles are then derived analytically for both layers. The refraction integral is used to find the atmospheric refraction and dispersion. 

Figure~\ref{fig:dispersionModels} also shows the SLALIB implementation of the \citeauthor{H&S1985} model. We surmise that neglecting the temperature inversion at higher altitudes results in small inaccuracies in the dispersion calculation, compared to the other spherical atmosphere models discussed earlier. 

More noteworthy, however, is that \citeauthor{H&S1985} use the outdated \cite{BS1939} equation for the refractive index. If we directly compare this to the Ciddor--Cassini model, using SLALIB, for an observation in \textit{H}-band at a zenith angle of 30\degr, we find a discrepancy of 4.1 mas in refraction and of 0.18 mas in dispersion. This reinforces the idea that selecting the right refractivity model is significantly more important than the choice of atmospheric model.

\subsection{Impact of local weather along the line of sight}
As a light ray travels through the atmosphere towards the telescope, it will encounter conditions deviating from the temperature and pressure profiles directly above the observatory. Most publically available three dimensional data sets either lack the spatial resolution or the altitudinal extension to investigate in detail the impact of such local weather on the atmospheric dispersion. Therefore, we have constructed a longitudinally extended atmosphere based on the U.S. Standard Atmosphere. Normally distributed perturbations of pressure and temperature are applied at altitudes between 1-2, 7-10, 25-30 and 50-60 km. The standard deviations at these points are arbitrarily chosen to be 1000, 500, 250 and 4 hPa for the atmospheric pressure and 5, 5, 10 and 15 K, for the temperature. Different perturbations are applied every 2 kilometers in longitudinal direction. Finally, we linearly interpolate the data to a resolution of 500 meters, in both altitude and longitude.

The differential between the refraction and dispersion as calculated from the atmosphere directly above the observer and those as calculated along the line of sight is generally of a small magnitude. Typical values are on the order of $10^{-2.3}$~mas for refraction and $10^{-5.5}$~mas for dispersion. A comparison using the publicly available NCEP North American Regional Reanalysis dataset \citep{Mesinger2006}, although of limited spatial resolution, gave results of similar magnitude.

\subsection{Differential dispersion as a result of optical properties}
The S-FPL51 and S-LAH71 glass used for the MICADO ADC prisms, offer a good representation of the dispersion of the atmosphere. It is not, however, a perfect immitation. When the dispersion correction is optimized for two wavelengths, the position of rays with different wavelength will not be optimal.

\begin{figure}
    \centering
    \includegraphics[width=\linewidth]{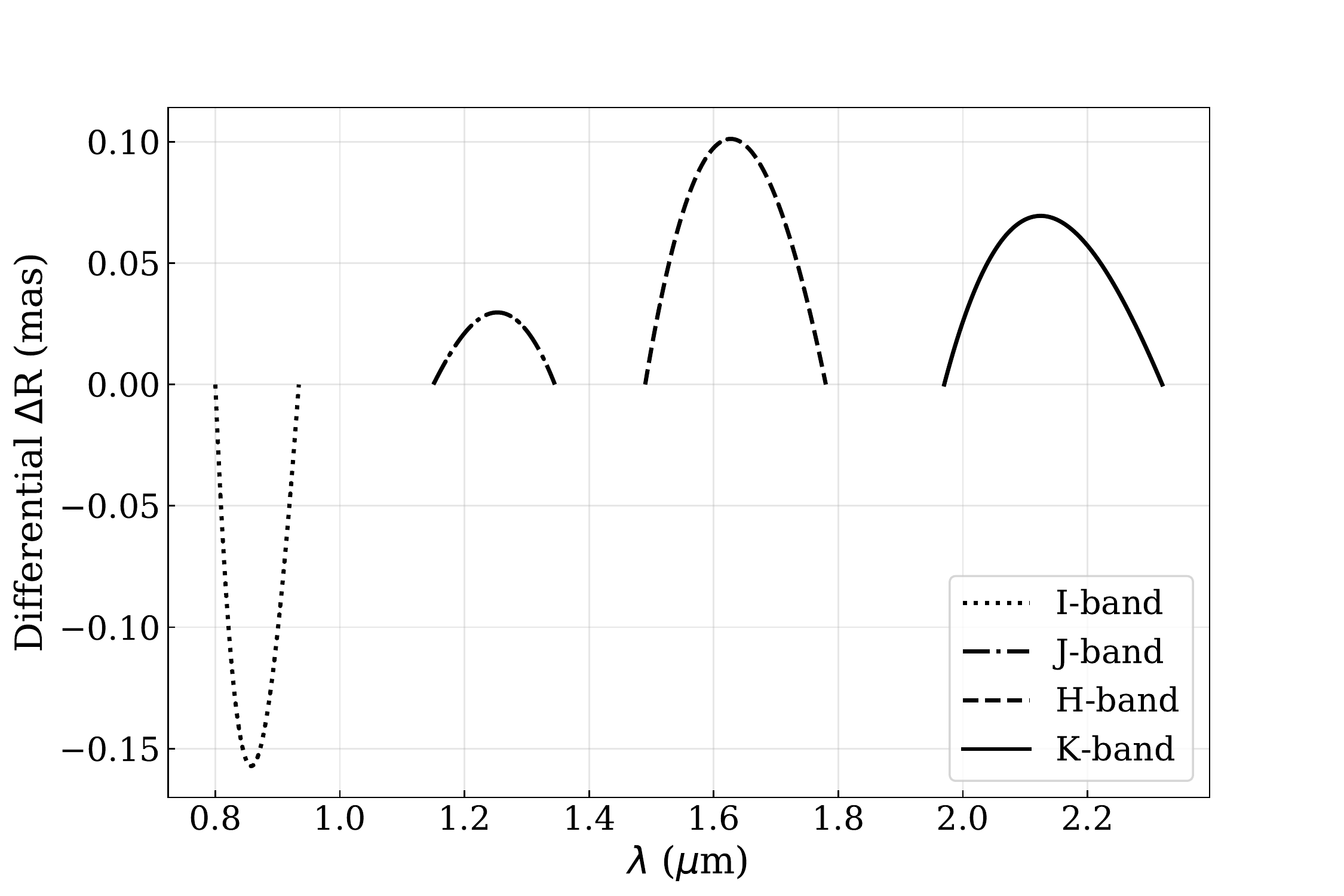}
    \caption{The difference in dispersion between the atmosphere and the current ADC design as a function of wavelength. The ADC position is optimized for the edge wavelengths of the respective band.}
    \label{fig:ADCcorrection}
\end{figure}

We illustrate this by using the Ciddor--Cassini atmospheric model as described in section~\ref{sec:AtmDispersion} and the ADC model as described in section~\ref{sec:ADC}. A zenith angle of 30$^{\circ}$ is assumed and for all broadband filters the ADC position is optimized for the edge wavelengths. The dispersion with respect to the shortest wavelength of the respective band is calculated for all wavelengths within the full filter band. The results are shown in Fig.~\ref{fig:ADCcorrection}. Unsurprisingly, the MICADO ADC has its optimum performance near the design wavelength of 1.35 \textmu m, though overall a differential dispersion up to 0.15 mas can be expected at this zenith angle. For larger zenith angles, the differential dispersion increases. Fortunately, these effects can be modeled well and are also constant over the full field for a given ADC position.

\subsection{Differential refraction as a result of different photometric color}
From our results above we might infer that the dispersion correction must also be dependent on the spectrum of the observed source. For any observed source that does not have a perfectly flat spectrum, the PSF will show a slight bias towards the wavelengths with higher intensity. We quantify the impact of photometric color on the astrometric accuracy by defining an effective wavelength and calculating the differential atmospheric refraction for two stars. The effective wavelength is defined as \cite{Gubler1998}
\begin{equation}
    \lambda_{\textnormal{eff}} = \frac{\int^{\lambda_{\textnormal{max}}}_{\lambda_{\textnormal{min}}}\lambda I(\lambda)d\lambda} {\int^{\lambda_{\textnormal{max}}}_{\lambda_{\textnormal{min}}} I(\lambda)d\lambda},
\end{equation}
where $\lambda_{\textnormal{min}}$ and $\lambda_{\textnormal{max}}$ are the boundaries of the wavelength range between which to integrate the intensity funcion, $I(\lambda)$. \citeauthor{Gubler1998} have shown that the above expression is equivalent, to within 1 \textmu as, to a similarly defined effective refraction, where the source spectrum and atmospheric and instrumental transmission have been taken into account.

Let us assume a blackbody spectrum for the observed source,
\begin{equation}
    % I(\lambda) = \frac{2\pi h c^2}{\lambda^5}\bigg( \frac{1}{e^{hc/\lambda k_b T} - 1} \bigg)
    I(\lambda) = \frac{2\pi h c^2}{\lambda^5}\left[\exp{\left(\frac{h c}{\lambda k_b T}\right)} - 1\right]^{-1}.
\end{equation}
Here $h$ is Planck's constant, $c$ is the speed of light, $k_b$ is Boltzmann's constant and $T$ is the effective temperature or color of the object. 

\begin{figure}
    \includegraphics[width=\linewidth]{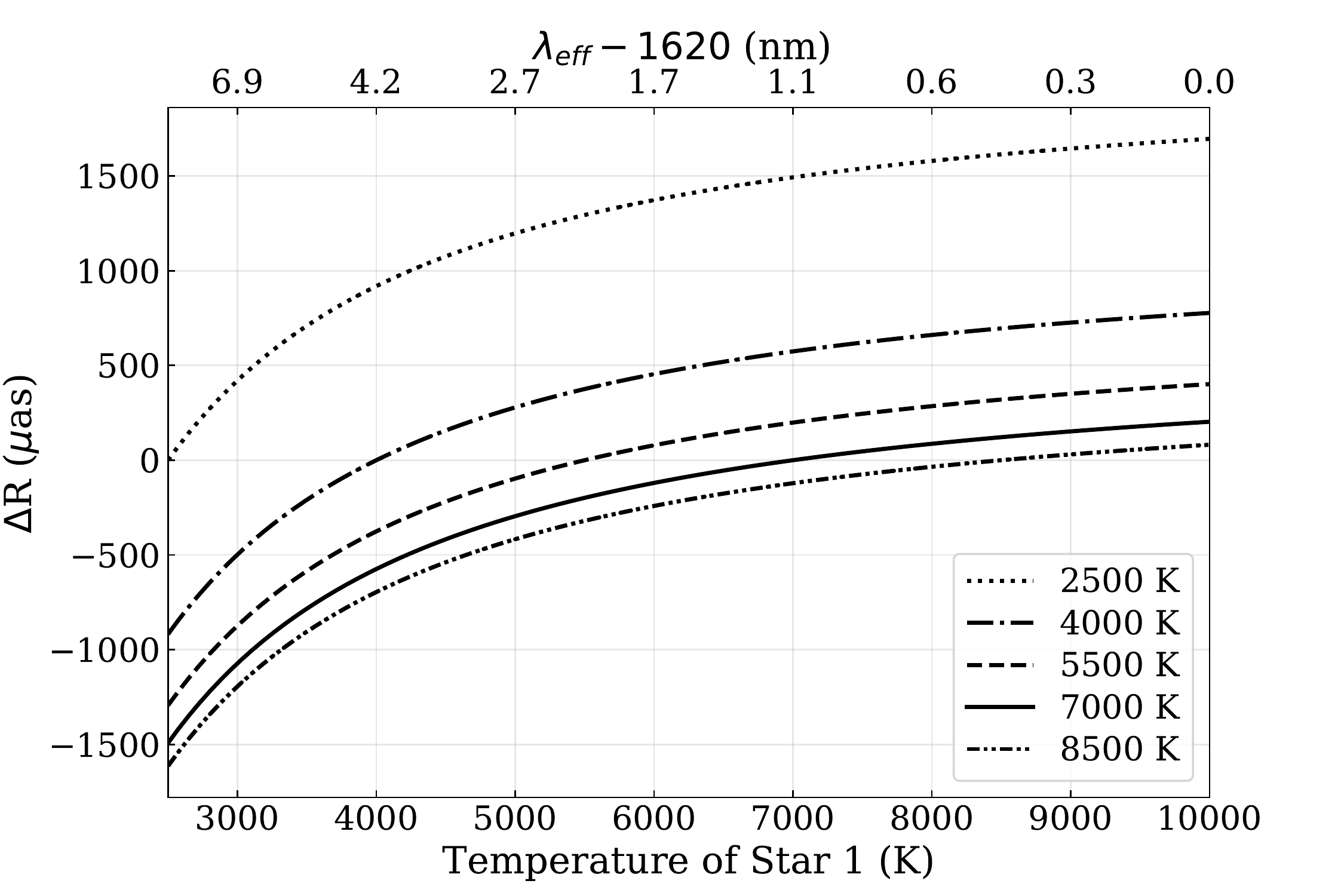}
    \caption{The differential refraction for two stars of different color for $z=60^{\circ}$ in \textit{H}-band, without the use of an ADC. The figure is adapted from Figure~1 of \protect\cite{Gubler1998}.}
    \label{fig:differential_refraction}
\end{figure}

Now, we compare the differential atmospheric refraction between two stars with different color at the same zenith angle. Figure~\ref{fig:differential_refraction} shows this in the extreme case that $z=60^{\circ}$, in \textit{H}-band. The differential refraction between the two stars is at most 1.5 mas, which is a small but not negligible fraction of the 54 mas dispersion that is observed over the whole bandpass at this zenith angle.

\begin{figure}
    \includegraphics[width=\linewidth]{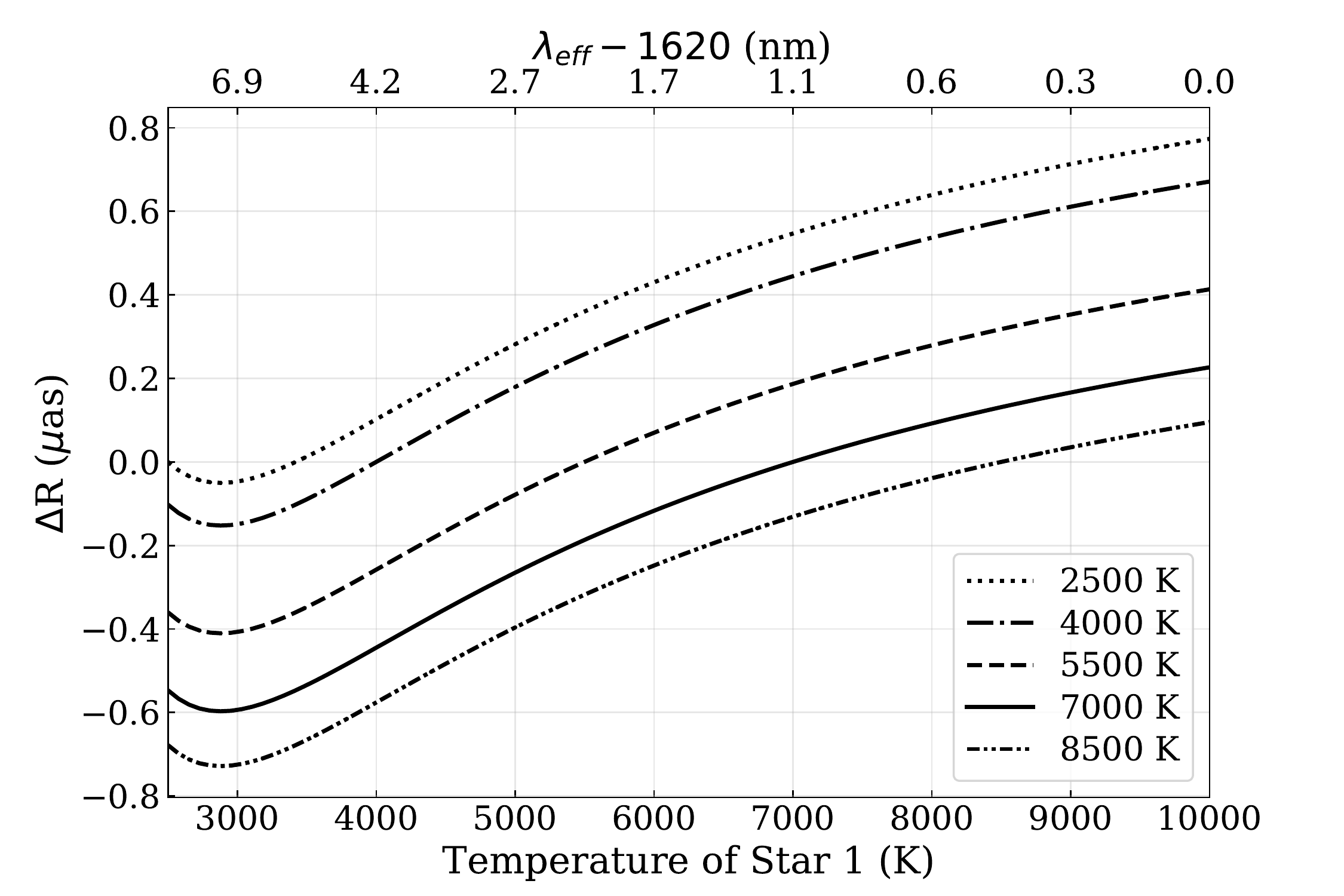}
    \caption{The differential refraction for two stars of different color for $z=60^{\circ}$ in \textit{H}-band, where an ADC is used and is optimized for the cut-on and cut-off wavelengths of the band.}
    \label{fig:differential_dispersion}
\end{figure}

With the addition of the MICADO ADC the contribution of this color dependence significantly decreases, as shown in Fig.~\ref{fig:differential_dispersion}. We've optimized the ADC position for the edge wavelengths of the filter passband. The differential refraction through the ADC is then subtracted from the differential atmospheric refraction for the effective wavelengths of the two stars. Now, the two stars show negligible differential refraction.


\subsection{Residual dispersion as a result of discretization of the dispersion correction}
\label{subsec:AtmDispersion/mech}
Using the equations of section~\ref{sec:mechTol} we calculate the necessary number of discrete steps to reach the 0.4 mas residual dispersion requirement, between $5^{\circ}$ and $45^{\circ}$ zenith angle. 

First, we use equation~(\ref{eq:filter_constant}) to compute the filter constant for \textit{H}-band manually and find a value $c_f$=$0.468$. We find that the maximum ADC rotation per milliarcsecond dispersion, at $z=5^{\circ}$, is 0.864$^{\circ}$. This corresponds to at least 1042 steps per 360\degr of prism rotation if a stepper motor is to be used.

Because of the assumptions of a plane-parallel atmosphere and of an ADC of which every optical interface is in the pupil plane, there will be a slight discrepancy in the obtained value of the filter constant in practice. To illustrate this, we fit equation~(\ref{eq:optPos}) to the optimum position of the MICADO optical model, using Zemax OpticStudio. There we obtain a value for the filter constant of $c_f$=$0.499$, resulting in at least 977 steps per full rotation of the prisms to meet the requirements. The present design has 1262.5 steps per rotation, which corresponds to a maximum of 0.31 mas residual dispersion and is therefore sufficient for both our analytical ADC model and the Zemax optical model.

For a posteriori calibrations the position of the ADC prisms are logged with a 16-bit absolute encoder. While the atmospheric dispersion must be reversed to within 0.4 mas level, for optimum astrometric performance the actual residual dispersion should be known to a higher accuracy. To achieve the astrometric goal of 10 \textmu as the position of the prisms must be known to about 31 arcsecond accuracy, or 4.1$\times 10^{4}$ steps over the full rotation. With the planned encoder on the prisms this will be sufficiently covered.

\section{Additional considerations}\label{sec:discussion}
There remain additional sources of residual dispersion likely to be present in the image plane of MICADO.

First of all, the adaptive optics system corrects the wavefront errors due to atmospheric turbulence with respect to some reference wavelength, typically outside the observed spectral range. The dispersive air then causes a wavelength dependent residual wavefront error. Several studies have assessed the impact of this chromatic path length error on the Strehl ratio, e.g. in \cite{Nakajima2006,Wallner1977,Devaney2008} and \cite{Jolissaint2010}. In addition, \cite{Nakajima2006} and \cite{Devaney2008} discuss the contribution of chromatic anisoplanatism to the degradation of Strehl ratio. They do not, however, quantify these effects in terms of angular size. The directions of such dispersion would, of course, be random and therefore correction by the ADC would not be possible. It none the less contributes to the uncertainty of the PSF centroid for astrometric calibrations.

Transmissive optics in the optical train may cause field dependent residual dispersion. Since the light beams originating from different positions in the field of view enter the optics with different angles of incidence, the refraction and dispersion may be slightly different. Except for the ADC, all transmissive optics in MICADO will be static during an observation. Any field dependent dispersion can therefore be easily modelled and calibrated.

In addition, at this point in time it is yet undecided how often the MICADO ADC will reorient itself during an observation. If the atmospheric conditions change significantly between two ADC optimizations, then the residual dispersion may be larger than expected. With good record keeping of the weather telemetry, this can be resolved after the observation, as is standard for telescopes that have implemented an ADC.

Finally, it is essential to accurately measure the atmospheric dispersion in the near infrared to verify the comparison of the dispersion models. This could be done using the trace map of a spectrometer with high enough resolution, in combination with altitudinal weather telemetry. Such an observation has been done by \cite{Skemer2009} in the mid infrared. Because of the limited sensitivity to the atmospheric dispersion model, these observations would primarily test the used refractivity model for typical conditions encountered at astronomical observing sites.

\section{Conclusions}\label{sec:conclusion}
The increased resolution and desired astrometric performance of upcoming imaging instruments on the next generation of extremely large telescopes required atmospheric dispersion to be studied in more detail than before. Specifically, for MICADO the residual chromatic dispersion on the detector is required to be smaller than 0.4 mas in \textit{H}-band to achieve the desired astrometric accuracy of 50 \textmu as. Therefore, we have investigated various contributors to the residual dispersion predicted to appear on the image plane. First, we compared seven atmospheric dispersion models using the same refractivity model and showed that, when a spherical atmosphere is assumed, the differential between these models is less than 10 \textmu as for $z<60$\degr. This implies that upcoming large observatories do not require new atmospheric models for the calculation of atmospheric dispersion. In contrast, a discrepancy of 0.18 mas in the atmospheric dispersion value was found by comparing the Ciddor-Cassini model to the equations of \cite{H&S1985}, used in the optical design software Zemax OpticStudio. Most of this discrepancy can be attributed to the refractivity model used in the respective atmospheric dispersion model. Therefore, we do not recommend using the equations of \citeauthor{H&S1985} for the control of an ADC on ELT instruments. Correspondingly, we would like to point out the limitations of Zemax OpticStudio when it comes to high-precision calculations of atmospheric refraction.

In addition, we performed error propagation on the Ciddor--Cassini dispersion model. Assuming a VLT Site Monitor and including an uncertainty of 1 nm in the wavelength, we found a systematic uncertainty of approximately 0.13 mas. We showed that the error in the cut-on and cut-off wavelengths of the passband dominates this uncertainty and should therefore be carefully defined or measured.

We have also investigated the impact of local atmospheric variations along the line of sight on the atmospheric dispersion calculation, using a full integration over the 1976 U.S. Standard Atmosphere. This proved to be such an insignificant effect that we conclude that it is not required for the correction of atmospheric dispersion to measure the full atmospheric profile, neither directly above the observatory, nor along the line of sight. Monitoring the local conditions at the observer remains important, as they do impact the amount of atmospheric dispersion.

Another investigated source of residual dispersion was the differential refraction between stars of different color. We showed that photometric color will not have to be taken into account for dispersion correction. When no ADC is present, the differential refraction can be up to 1.5 mas in \textit{H}-band. This reduces to less than 1 \textmu as if an ADC is included, even for large zenith distances.

Finally, we evaluated the optomechanical design of the MICADO ADC. The angular resolution required to reach a dispersion correction with less than 0.4 mas residual dispersion was calculated. At least 977 steps per rotation are needed to achieve the requirement between $z=5$\degr and $z=45$\degr. The present design will comply with this number. The differential dispersion due to the optical properties of the glass types used in the prisms causes a residual dispersion up to 0.15 mas for a modest $z=30$\degr. 

Overall, the limited positioning accuracy will be the largest contributor to the residual dispersion, with a maximum contribution of 0.31 mas before readjustment. Accurate measurements of the refractive indices of the S-FPL51 and S-LAH71 glass and accurate measurements of the rotation of the prisms during operations will allow the optomechanical contributors to be well known. The other discussed effects are less well known, since they depend on more variables. Fortunately, they are generally of smaller magnitude. 

In conclusion, we believe that the chromatic dispersion in the instrument is now well enough understood that it will not prevent MICADO from realising its full astrometric potential.

\section*{Acknowledgements}

We'd like to express our gratitude to Eline Tolstoy, Bayu Jayawardhana, Ramon Navarro and Annemieke Janssen for their discussions and thoughtful feedback during this research, to Michael Hartl for providing the optical design of the MICADO instrument and to the MICADO consortium for its support. We'd also like to thank the reviewer his or her useful feedback.

The NARR data are provided by the NOAA/OAR/ESRL PSD, Boulder, Colorado, USA. It can be retrieved from their website at \url{https://www.esrl.noaa.gov/psd/}. [Accessed: 27-06-2019].

Our Python code is available from the first author or at: \url{https://gitlab.astro-wise.org/micado/atmosphericmodel}

%%%%%%%%%%%%%%%%%%%%%%%%%%%%%%%%%%%%%%%%%%%%%%%%%%



\bibliographystyle{mnras}
\bibliography{Bibliography} % if your bibtex file is called example.bib

\begin{thebibliography}{}
\makeatletter
\relax
\def\mn@urlcharsother{\let\do\@makeother \do\$\do\&\do\#\do\^\do\_\do\%\do\~}
\def\mn@doi{\begingroup\mn@urlcharsother \@ifnextchar [ {\mn@doi@}
  {\mn@doi@[]}}
\def\mn@doi@[#1]#2{\def\@tempa{#1}\ifx\@tempa\@empty \href
  {http://dx.doi.org/#2} {doi:#2}\else \href {http://dx.doi.org/#2} {#1}\fi
  \endgroup}
\def\mn@eprint#1#2{\mn@eprint@#1:#2::\@nil}
\def\mn@eprint@arXiv#1{\href {http://arxiv.org/abs/#1} {{\tt arXiv:#1}}}
\def\mn@eprint@dblp#1{\href {http://dblp.uni-trier.de/rec/bibtex/#1.xml}
  {dblp:#1}}
\def\mn@eprint@#1:#2:#3:#4\@nil{\def\@tempa {#1}\def\@tempb {#2}\def\@tempc
  {#3}\ifx \@tempc \@empty \let \@tempc \@tempb \let \@tempb \@tempa \fi \ifx
  \@tempb \@empty \def\@tempb {arXiv}\fi \@ifundefined
  {mn@eprint@\@tempb}{\@tempb:\@tempc}{\expandafter \expandafter \csname
  mn@eprint@\@tempb\endcsname \expandafter{\@tempc}}}

\bibitem[\protect\citeauthoryear{{Auer} \& {Standish}}{{Auer} \&
  {Standish}}{2000}]{Auer2000}
{Auer} L.~H.,  {Standish} E.~M.,  2000, \mn@doi [Astronomical Journal]
  {10.1086/301325}, \href
  {https://ui.adsabs.harvard.edu/abs/2000AJ....119.2472A} {119, 2472}

\bibitem[\protect\citeauthoryear{{Barrell} \& {Sears}}{{Barrell} \&
  {Sears}}{1939}]{BS1939}
{Barrell} H.,  {Sears} J.~E.,  1939, \mn@doi [Philosophical Transactions of the
  Royal Society of London Series A] {10.1098/rsta.1939.0004}, \href
  {http://adsabs.harvard.edu/abs/1939RSPTA.238....1B} {238, 1}

\bibitem[\protect\citeauthoryear{{Birch} \& {Downs}}{{Birch} \&
  {Downs}}{1993}]{Birch1993}
{Birch} K.~P.,  {Downs} M.~J.,  1993, \mn@doi [Metrologia]
  {10.1088/0026-1394/30/3/004}, \href
  {http://adsabs.harvard.edu/abs/1993Metro..30..155B} {30, 155}

\bibitem[\protect\citeauthoryear{{B{\"o}nsch} \& {Potulski}}{{B{\"o}nsch} \&
  {Potulski}}{1998}]{Bonsch1998}
{B{\"o}nsch} G.,  {Potulski} E.,  1998, \mn@doi [Metrologia]
  {10.1088/0026-1394/35/2/8}, \href
  {http://adsabs.harvard.edu/abs/1998Metro..35..133B} {35, 133}

\bibitem[\protect\citeauthoryear{{Ciddor}}{{Ciddor}}{1996}]{Ciddor1996}
{Ciddor} P.~E.,  1996, \mn@doi [Applied Optics] {10.1364/AO.35.001566}, \href
  {http://adsabs.harvard.edu/abs/1996ApOpt..35.1566C} {35, 1566}

\bibitem[\protect\citeauthoryear{{Corbard}, {Ikhlef}, {Morand}, {Meftah}  \&
  {Renaud}}{{Corbard} et~al.}{2019}]{Corbard2019}
{Corbard} T.,  {Ikhlef} R.,  {Morand} F.,  {Meftah} M.,   {Renaud} C.,  2019,
  \mn@doi [MNRAS] {10.1093/mnras/sty3391}, \href
  {http://adsabs.harvard.edu/abs/2019MNRAS.483.3865C} {483, 3865}

\bibitem[\protect\citeauthoryear{{Danjon}}{{Danjon}}{1980}]{Danjon1980}
{Danjon} A.,  1980, {Astronomie generale. Astronomie spherique et elements de
  mecanique celeste}.
Paris : Albert Blanchard, 1980, 2nd ed.

\bibitem[\protect\citeauthoryear{{Davies} et~al.,}{{Davies}
  et~al.}{2016}]{Davies2016}
{Davies} R.,  et~al., 2016, ] {10.1117/12.2233047}, \href
  {https://ui.adsabs.harvard.edu/abs/2016SPIE.9908E..1ZD} {9908, 99081Z}

\bibitem[\protect\citeauthoryear{{Davis}}{{Davis}}{1992}]{Davis1992}
{Davis} R.~S.,  1992, \mn@doi [Metrologia] {10.1088/0026-1394/29/1/008}, \href
  {http://adsabs.harvard.edu/abs/1992Metro..29...67D} {29, 67}

\bibitem[\protect\citeauthoryear{{Devaney}, {Goncharov}  \& {Dainty}}{{Devaney}
  et~al.}{2008}]{Devaney2008}
{Devaney} N.,  {Goncharov} A.~V.,   {Dainty} J.~C.,  2008, \mn@doi [\ao]
  {10.1364/AO.47.001072}, \href
  {https://ui.adsabs.harvard.edu/abs/2008ApOpt..47.1072D} {47, 1072}

\bibitem[\protect\citeauthoryear{ESO}{ESO}{2011}]{ESOconstrprop}
ESO 2011, {E-ELT Construction Proposal},
  \url{https://www.eso.org/public/archives/books/pdf/book_0046.pdf}

\bibitem[\protect\citeauthoryear{{Egner} et~al.,}{{Egner}
  et~al.}{2010}]{Egner2010}
{Egner} S.,  et~al., 2010, in Adaptive Optics Systems II. p. 77364V,
  \mn@doi{10.1117/12.856579}

\bibitem[\protect\citeauthoryear{{Ellerbroek}}{{Ellerbroek}}{2013}]{Ellerbroek2013}
{Ellerbroek} B.,  2013, \mn@doi [\aap] {10.1051/0004-6361/201321092}, \href
  {https://ui.adsabs.harvard.edu/abs/2013A&A...552A..41E} {552, A41}

\bibitem[\protect\citeauthoryear{{Fletcher}}{{Fletcher}}{1931}]{Fletcher1931}
{Fletcher} A.,  1931, \mn@doi [MNRAS] {10.1093/mnras/91.5.559}, \href
  {http://adsabs.harvard.edu/abs/1931MNRAS..91..559F} {91, 559}

\bibitem[\protect\citeauthoryear{{Gubler} \& {Tytler}}{{Gubler} \&
  {Tytler}}{1998}]{Gubler1998}
{Gubler} J.,  {Tytler} D.,  1998, \mn@doi [PASP] {10.1086/316172}, \href
  {http://adsabs.harvard.edu/abs/1998PASP..110..738G} {110, 738}

\bibitem[\protect\citeauthoryear{{Hagen} \& {Tkaczyk}}{{Hagen} \&
  {Tkaczyk}}{2011}]{Hagen2011}
{Hagen} N.,  {Tkaczyk} T.~S.,  2011, \mn@doi [\ao] {10.1364/AO.50.004998},
  \href {https://ui.adsabs.harvard.edu/abs/2011ApOpt..50.4998H} {50, 4998}

\bibitem[\protect\citeauthoryear{{Heiskanen} \& {Moritz}}{{Heiskanen} \&
  {Moritz}}{1967}]{HeiskanenMoritz1967}
{Heiskanen} W.~A.,  {Moritz} H.,  1967, \mn@doi [Bulletin Geodesique]
  {10.1007/BF02525647}, \href
  {https://ui.adsabs.harvard.edu/abs/1967BGeod..41..491H} {41, 491}

\bibitem[\protect\citeauthoryear{Hohenkerk \& Sinclair}{Hohenkerk \&
  Sinclair}{1985}]{H&S1985}
Hohenkerk C.,  Sinclair A.,  1985, The Computation of an Angular Atmospheric
  Refraction at Large Zenith Angles.
NAO Technical note, H.M. Stationery Office

\bibitem[\protect\citeauthoryear{{Johns} et~al.,}{{Johns}
  et~al.}{2012}]{Johns2012}
{Johns} M.,  et~al., 2012, in Ground-based and Airborne Telescopes IV. p.
  84441H, \mn@doi{10.1117/12.926716}

\bibitem[\protect\citeauthoryear{{Jolissaint} \& {Kendrew}}{{Jolissaint} \&
  {Kendrew}}{2010}]{Jolissaint2010}
{Jolissaint} L.,  {Kendrew} S.,  2010, in Adaptative Optics for Extremely Large
  Telescopes. p. 05021, \mn@doi{10.1051/ao4elt/201005021}

\bibitem[\protect\citeauthoryear{Kragh}{Kragh}{2018}]{Kragh2018}
Kragh H.,  2018, \mn@doi [Substantia] {10.13128/Substantia-56}, 2, 7

\bibitem[\protect\citeauthoryear{{Mangum} \& {Wallace}}{{Mangum} \&
  {Wallace}}{2015}]{Mangum2015}
{Mangum} J.~G.,  {Wallace} P.,  2015, \mn@doi [PASP] {10.1086/679582}, \href
  {https://ui.adsabs.harvard.edu/abs/2015PASP..127...74M} {127, 74}

\bibitem[\protect\citeauthoryear{{Massari} et~al.,}{{Massari}
  et~al.}{2016}]{Massari2016}
{Massari} D.,  et~al., 2016, ] {10.1117/12.2232478}, \href
  {https://ui.adsabs.harvard.edu/abs/2016SPIE.9909E..1GM} {9909, 99091G}

\bibitem[\protect\citeauthoryear{{Mathar}}{{Mathar}}{2007}]{Mathar2007}
{Mathar} R.~J.,  2007, \mn@doi [Journal of Optics A: Pure and Applied Optics]
  {10.1088/1464-4258/9/5/008}, \href
  {https://ui.adsabs.harvard.edu/abs/2007JOptA...9..470M} {9, 470}

\bibitem[\protect\citeauthoryear{{Mesinger} et~al.,}{{Mesinger}
  et~al.}{2006}]{Mesinger2006}
{Mesinger} F.,  et~al., 2006, \mn@doi [Bulletin of the American Meteorological
  Society] {10.1175/BAMS-87-3-343}, \href
  {https://ui.adsabs.harvard.edu/abs/2006BAMS...87..343M} {87, 343}

\bibitem[\protect\citeauthoryear{{Moritz}}{{Moritz}}{2000}]{GRS80}
{Moritz} H.,  2000, \mn@doi [Journal of Geodesy] {10.1007/s001900050278}, \href
  {https://ui.adsabs.harvard.edu/abs/2000JGeod..74..128M} {74, 128}

\bibitem[\protect\citeauthoryear{{NOAA}}{{NOAA}}{1976}]{1976USSA}
{NOAA} 1976, Technical report, {U.S. standard atmosphere, 1976}.
NASA

\bibitem[\protect\citeauthoryear{{Nakajima}}{{Nakajima}}{2006}]{Nakajima2006}
{Nakajima} T.,  2006, \mn@doi [\apj] {10.1086/508418}, \href
  {https://ui.adsabs.harvard.edu/abs/2006ApJ...652.1782N} {652, 1782}

\bibitem[\protect\citeauthoryear{{Nurzia}}{{Nurzia}}{2018}]{Nurzia2018}
{Nurzia} V.,  2018, in Proc.~SPIE. p. 1070403, \mn@doi{10.1117/12.2312049}

\bibitem[\protect\citeauthoryear{Oriani}{Oriani}{1787}]{Oriani1787}
Oriani B.,  1787, Ephemerides astronomicae anni 1788: Appendix ad ephemerides
  Anni 1788

\bibitem[\protect\citeauthoryear{{Picard}, {Davis}, {Gl{\"a}ser}  \&
  {Fujii}}{{Picard} et~al.}{2008}]{Picard2008}
{Picard} A.,  {Davis} R.~S.,  {Gl{\"a}ser} M.,   {Fujii} K.,  2008, \mn@doi
  [Metrologia] {10.1088/0026-1394/45/2/004}, \href
  {http://adsabs.harvard.edu/abs/2008Metro..45..149P} {45, 149}

\bibitem[\protect\citeauthoryear{{Pott} \& {Davies}}{{Pott} \&
  {Davies}}{2018}]{Pott2018}
{Pott} J.-U.,  {Davies} R.,  2018, MICADO astrometry system architecture \-
  design, simulations, and error budget, MICADO PDR (internal review doc.)

\bibitem[\protect\citeauthoryear{{Sanders}}{{Sanders}}{2014}]{Sanders2014}
{Sanders} G.,  2014, in Thirty Meter Telescope Science Forum. p.~60

\bibitem[\protect\citeauthoryear{{Sandrock}, {Amestica}  \&
  {Sarazin}}{{Sandrock} et~al.}{1999}]{VLTASM}
{Sandrock} S.,  {Amestica} R.,   {Sarazin} M.,  1999, VLT Astronomical Site
  Monitor ASM Data User Manual, ESO

\bibitem[\protect\citeauthoryear{{Schoeck}, {Do}, {Ellerbroek}, {Herriot},
  {Meyer}, {Suzuki}, {Wang}  \& {Yelda}}{{Schoeck} et~al.}{2013}]{Schoeck2013}
{Schoeck} M.,  {Do} T.,  {Ellerbroek} B.,  {Herriot} G.,  {Meyer} L.,  {Suzuki}
  R.,  {Wang} L.,   {Yelda} S.,  2013, in {Esposito} S.,  {Fini} L.,  eds,
  Proceedings of the Third AO4ELT Conference. p.~77,
  \mn@doi{10.12839/AO4ELT3.13356}

\bibitem[\protect\citeauthoryear{{Skemer} et~al.,}{{Skemer}
  et~al.}{2009}]{Skemer2009}
{Skemer} A.~J.,  et~al., 2009, \mn@doi [PASP] {10.1086/605312}, \href
  {http://adsabs.harvard.edu/abs/2009PASP..121..897S} {121, 897}

\bibitem[\protect\citeauthoryear{{Span{\`o}}}{{Span{\`o}}}{2014}]{Spano2014}
{Span{\`o}} P.,  2014, in Advances in Optical and Mechanical Technologies for
  Telescopes and Instrumentation. p. 915157, \mn@doi{10.1117/12.2057072}

\bibitem[\protect\citeauthoryear{{Trippe}, {Davies}, {Eisenhauer}, {Schreiber},
  {Fritz}  \& {Genzel}}{{Trippe} et~al.}{2010}]{Trippe2010}
{Trippe} S.,  {Davies} R.,  {Eisenhauer} F.,  {Schreiber} N.~M.~F.,  {Fritz}
  T.~K.,   {Genzel} R.,  2010, \mn@doi [\mnras]
  {10.1111/j.1365-2966.2009.15940.x}, \href
  {https://ui.adsabs.harvard.edu/abs/2010MNRAS.402.1126T} {402, 1126}

\bibitem[\protect\citeauthoryear{{Wallace}}{{Wallace}}{2005}]{SLALIB}
{Wallace} P.~T.,  2005, Starlink User Note, \href
  {http://adsabs.harvard.edu/abs/2005StaUN..67.....W} {67}

\bibitem[\protect\citeauthoryear{{Wallner}}{{Wallner}}{1977}]{Wallner1977}
{Wallner} E.~P.,  1977, Journal of the Optical Society of America (1917-1983),
  \href {https://ui.adsabs.harvard.edu/abs/1977JOSA...67..407W} {67, 407}

\bibitem[\protect\citeauthoryear{{Young}}{{Young}}{2006}]{Young2006}
{Young} A.~T.,  2006, The Observatory, \href
  {http://adsabs.harvard.edu/abs/2006Obs...126...82Y} {126, 82}

\makeatother
\end{thebibliography}

%%%%%%%%%%%%%%%%%%%%%%%%%%%%%%%%%%%%%%%%%%%%%%%%%%

%%%%%%%%%%%%%%%%% APPENDICES %%%%%%%%%%%%%%%%%%%%%

\appendix

\section{Uncertainties of the Ciddor--Cassini model}\label{appendix:uncertainties}
In addition to the discussion in section~\ref{subsec:results_and_errors} on the error propagation of the Ciddor--Cassini dispersion model, we have explored the parameter space of the uncertainties to find the most sensitive components of the model.

\begin{figure*}
\centering
\subfloat[\label{fig:error_in_P}]{\includegraphics[width=0.33\textwidth]{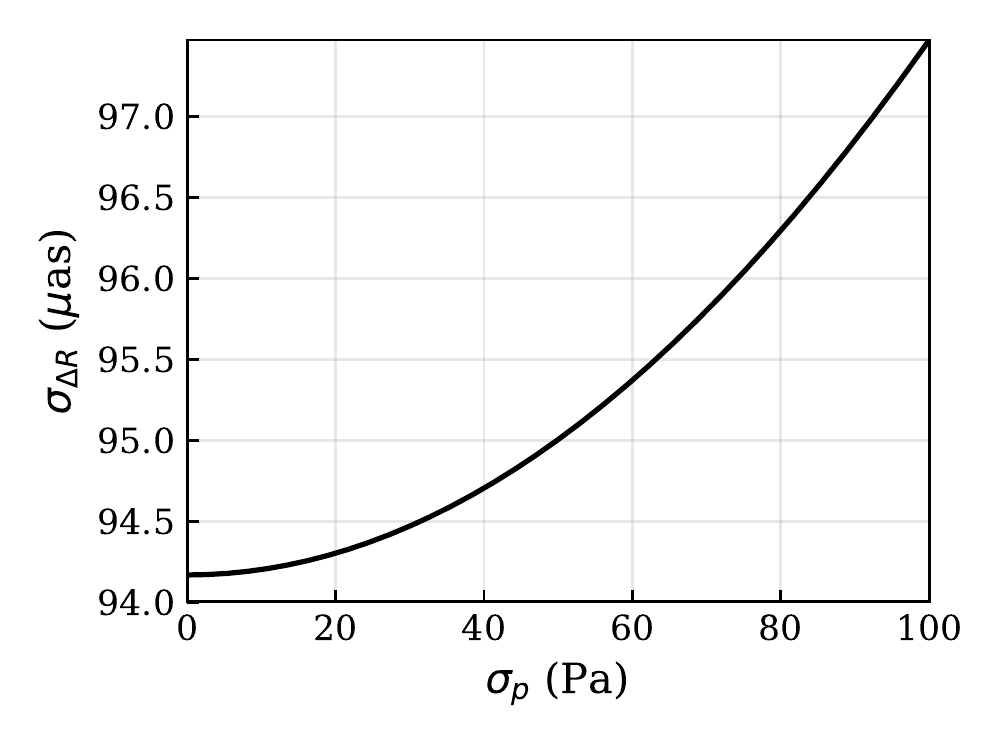}}\hfill
\subfloat[\label{fig:error_in_T}] {\includegraphics[width=0.33\textwidth]{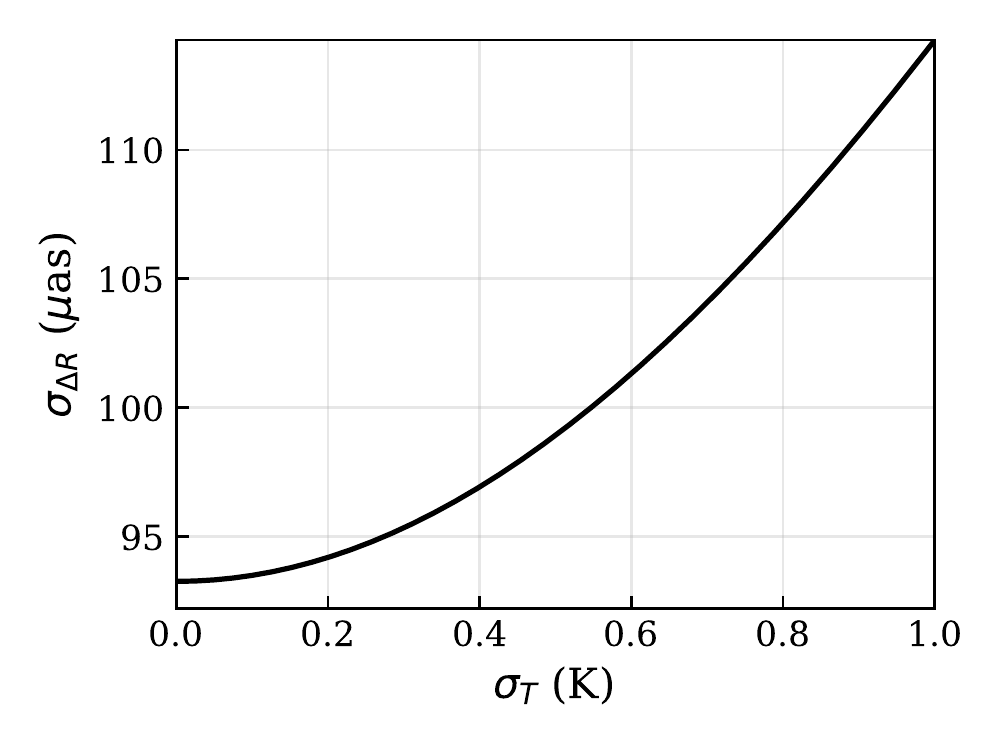}}\hfill
\subfloat[\label{fig:error_in_H}]{\includegraphics[width=0.33\textwidth]{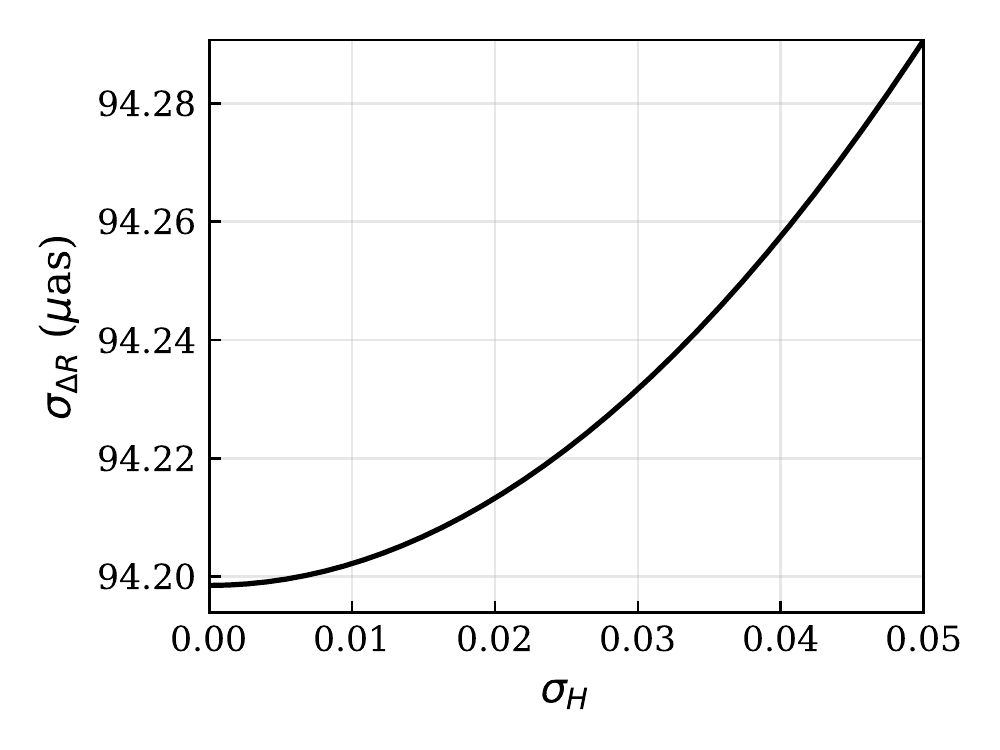}}\hfill\\
\subfloat[\label{fig:error_in_xc}]{\includegraphics[width=0.33\textwidth]{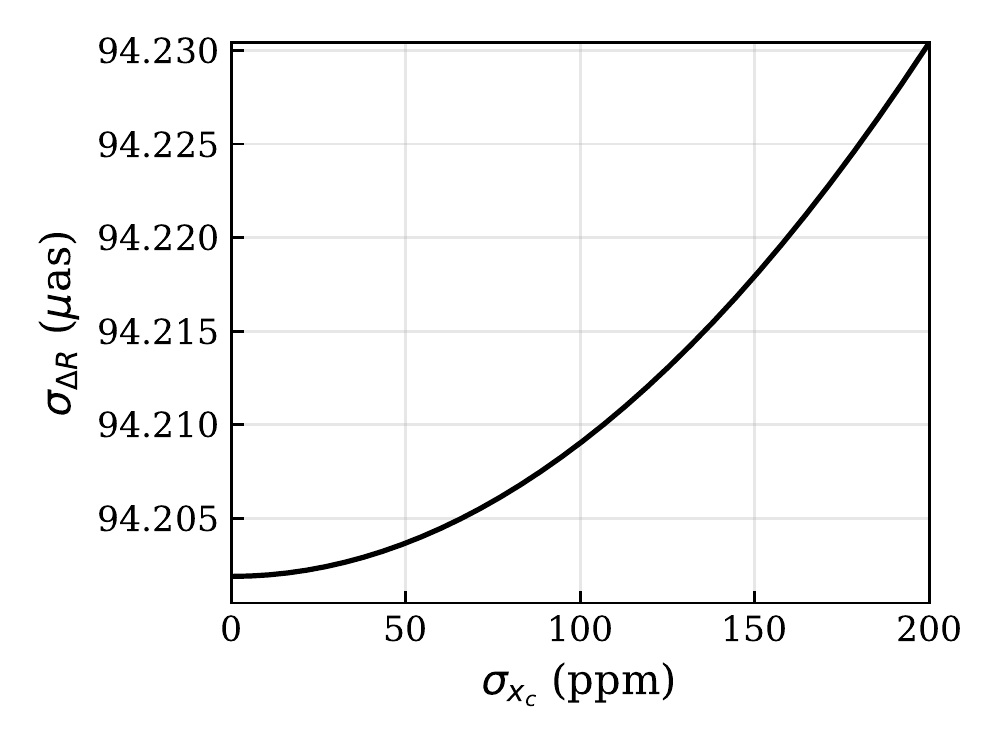}}\hfill
\subfloat[\label{fig:error_in_wavelength}] {\includegraphics[width=0.33\textwidth]{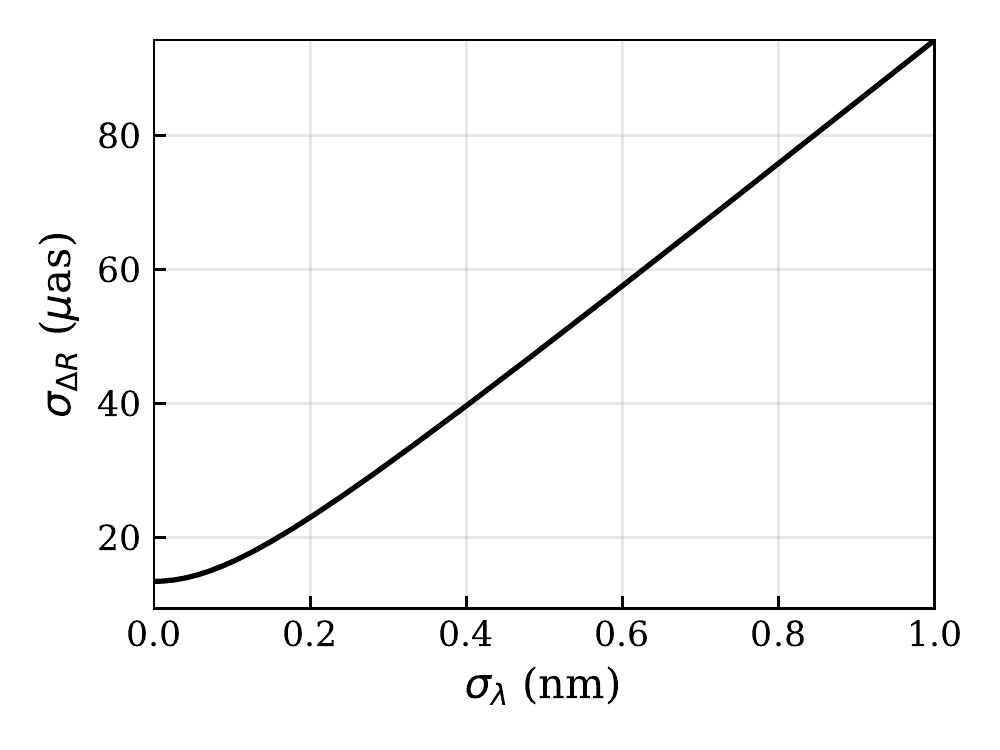}}\hfill
\subfloat[\label{fig:error_in_z}]{\includegraphics[width=0.33\textwidth]{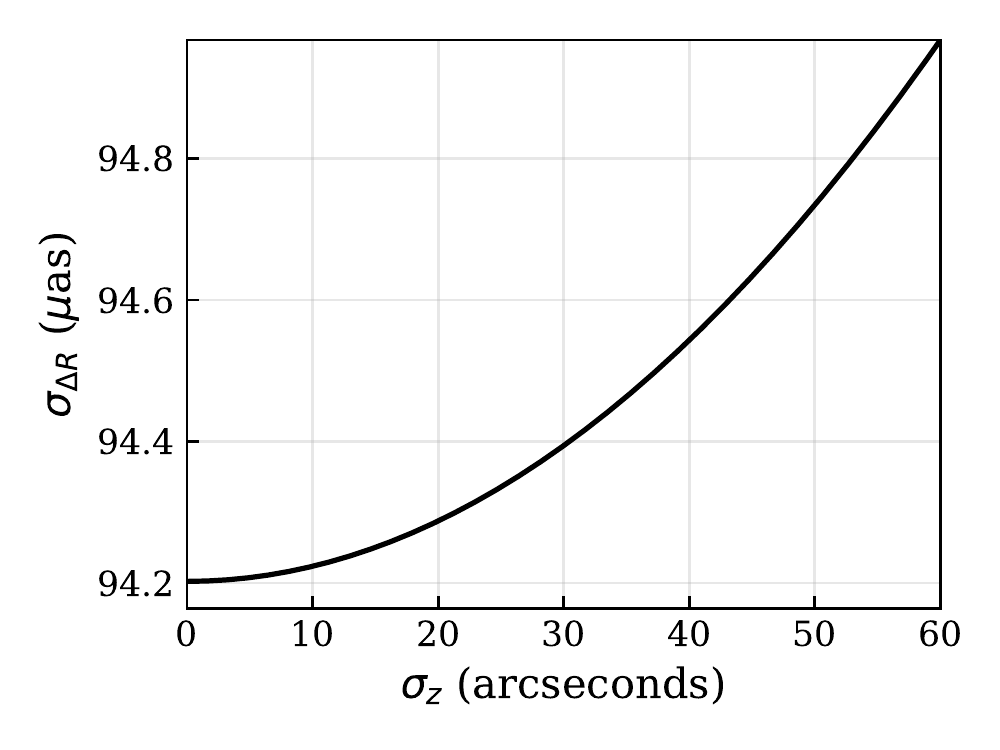}}\hfill\\
\caption{The uncertainty in the atmospheric dispersion, $\sigma_{\Delta R}$, as a function of the measurement uncertainty in (a) pressure, (b) temperature, (c) relative humidity, (d) CO$_2$ density, (e) wavelength and (f) zenith angle. Except for the respective uncertainty variable, the following uncertainties are assumed: $\sigma_p=10$ Pa, $\sigma_T=0.2^{\circ}$C, $\sigma_H=0.01$, $\sigma_{x_c}=20$ ppm, $\sigma_{\lambda}=0.001$ \textmu m and $\sigma_z=0^{\circ}$.}
\label{fig:errors}
\end{figure*}

We have taken the typical measurement uncertainties from the VLT Site Monitor \citep{VLTASM}. Explicitly, $\sigma_p=10$ Pa, $\sigma_T=0.2^{\circ}$C, $\sigma_H=0.01$ and $\sigma_{x_c}=20$ ppm. We assume $\sigma_z=0^{\circ}$. To explore the parameter space, we have varied any single variable uncertainty of the parameters included in our analysis (i.e. $\sigma_p$, $\sigma_T$, $\sigma_h$, $\sigma_{x_c}$, $\sigma_{\lambda}$ and $\sigma_z$), while keeping all the others at their typical values. The results are shown in Figure~\ref{fig:errors}.

All the uncertainties behave linearly and the curvature in the figures is only present before the given uncertainty is large enough to dominate over the other parameters.

% Don't change these lines
\bsp	% typesetting comment
\label{lastpage}
\end{document}